\documentclass[epj]{mysvjour}
\usepackage{graphics}
\usepackage{epsfig}
\usepackage{psfrag}

\newcommand{\bbms}{$\mathrm{B_s}\!-\!\ov{\mathrm{B}}{}_\mathrm{s}\,$\ mixing}

\newcommand{\lt}{\left}
\newcommand{\rt}{\right}
\newcommand{\no}{\nonumber}
\newcommand{\nn}{\nonumber \\}
\newcommand{\ov}[1]{\overline{#1}}
\newcommand{\eq}[1]{(\ref{#1})}
\newcommand{\be}{\begin{equation}}
\newcommand{\ee}{\end{equation}}
\newcommand{\bea}{\begin{eqnarray}}
\newcommand{\eea}{\end{eqnarray}}

\newcommand{\tg}{{\tilde g}}

\newcommand{\mtpole}{{m_t^{\mbox{\rm \scriptsize pole}}}}
\newcommand{\tq}{{\tilde q}}
\newcommand{\stp}{{\tilde t}}
\newcommand{\gev}{\ensuremath{\,\mbox{GeV}}}
\newcommand{\mgut}{{M_{\mbox{\rm \scriptsize GUT}}}}
\newcommand{\mten}{{M_{10}}}
\newcommand{\mpl}{{M_{\mbox{\rm \scriptsize Pl}}}}
\newcommand{\bra}[1]{\langle \, #1 \, | }
\newcommand{\ket}[1]{| \, #1 \, \rangle }

\begin{document}
\title{\vspace{-1.5cm}\\ 
\boldmath
\bbms\ in an SO(10) SUSY GUT model\thanks{talk at 
\emph{HEP2003 Europhysics Conference}, 17-23 Jul 2003, Aachen, Germany}
\unboldmath}
\author{Sebastian J\"ager\inst{1}%
\thanks{\emph{Present address:} Institut f\"ur Theoretische Physik E, 
RWTH Aachen, 52056 Aachen, Germany.}
 \and Ulrich Nierste\inst{2}%
\thanks{speaker}}
%
%
\institute{Physik-Department T31, 
Technische Universit\"at M\"unchen, 85748 Garching, Germany.
\and 
    Fermi National Accelerator Laboratory, Batavia,
        IL 60510-500, USA. 
    }
\date{Received: 7 Dec 2003}
%
\abstract{We perform a renormalisation group analysis of 
the SO(10) model proposed by Chang, Masiero and Murayama, which links
the large atmospheric neutrino mixing angle to loop-induced
transitions between right-handed $b$ and $s$ quarks. We compute the
impact on \bbms\ and find that the mass difference in the $\mathrm{B_s}$ system
can exceed its Standard Model value by a factor of 16.
} 
\maketitle
\section{Introduction}
\label{intro}
Grand Unified Theories combine quarks and leptons into symmetry
multiplets. This opens the possibility to find links between the
flavour structures of both sectors. Chang, Masiero and Murayama (CMM)
have proposed an interesting supersymmetric SO(10) model in which the
large $\nu_\mu$--$\nu_\tau$ mixing angle can affect transitions
between right-handed $b$ and $s$ quarks through supersymmetric loop
diagrams
\cite{cmm}. The SO(10)-symmetric superpotential has the form
\begin{eqnarray}
W_{10} &=& \frac{1}{2} \mathbf{16}^T Y^U \mathbf{16} \, \mathbf{10}_\mathbf{H} 
  \; + \; 
   \frac{1}{\mpl} \frac{1}{2} \mathbf{16}^T \tilde{Y}^D  \mathbf{16}
     \, \mathbf{10_H^\prime} \mathbf{45_H} \nn
&& + \frac{1}{\mpl} \frac{1}{2}  \mathbf{16}^T Y^M \mathbf{16}  
   \, \mathbf{\ov{16}_H} \mathbf{\ov{16}_H}. \label{w10}
\end{eqnarray} 
Here $\mathbf{16}$ is the usual spinor comprising the matter
superfields and the other fields are Higgs superfields in the
indicated representations. The Yukawa coupling $Y^U$ is a symmetric
$3\times 3$ matrix in generation space containing the large top Yukawa
coupling. The two dimension-5 terms involve the Planck mass $\mpl$
and further Higgs fields in the indicated representations. The last
term in \eq{w10} generates small neutrino masses via the standard
see-saw mechanism, once $\mathbf{\ov{16}_H}$ acquires an
SO(10)-breaking VEV. From $m_t \gg m_c$ we observe a large hierarchy
in $Y^U$, which must be largely compensated in $Y^M$ in order to
explain the observed pattern of neutrino masses. This is achieved in a
natural way by invoking flavour symmetries, the simplest of which
render $Y^U$ and $Y^M$ simultaneously diagonal. In this basis the
remaining Yukawa matrix $\tilde{Y}^D$ has the form
\begin{eqnarray}
\tilde{Y}^D &=& V_{\rm CKM}^* \, 
\lt(\begin{array}{ccc}
\tilde y_d&0&0\\0& \tilde y_s&0\\ 0&0&\tilde y_b\end{array}\rt)
\, U_{\rm PMNS} . \label{ydd}
\end{eqnarray}
Here $V_{\rm CKM}$ and $U_{\rm PMNS}$ are the CKM and PMNS matrices
encoding flavour mixing in the quark and lepton sectors \cite{ckm} and
certain diagonal phase matrices have been omitted in \eq{ydd}. From
\eq{ydd} one realizes that the flavour structure of $\tilde{Y}^D$ is
neither symmetric nor anti-symmetric. This is possible, because the
corresponding dimension-5 term in \eq{w10} transforms reducibly under
SO(10). At some scale $\mten$ between the Planck and GUT scales 
the $\mathbf{45_H}$ and $\mathbf{\ov{16}_H}$ acquire VEVs and SO(10)
is broken to SU(5). The SU(5) superpotential reads
\begin{eqnarray} 
W_5 &=& \frac{1}{2} \Psi^T Y^U \Psi \mathbf{5}_\mathbf{H} \, + \,  
	\Psi^T Y^D \Phi \, \mathbf{\ov{5}_H} \, +\, 
	\Phi^T  Y^\nu N \, \mathbf{5}_\mathbf{H} \nn 
&&+\, 
         \frac{1}{2}\frac{v_{\ov{16}}^2}{\mpl} \, 
         N^T Y^M N . \label{w5} 
\end{eqnarray}
The matter supermultiplets $\Psi$, $\Phi$ and $N$ are the usual
$\mathbf{10}$, $\mathbf{\ov 5}$ and $\mathbf{1}$ from the
decomposition of the $\mathbf{16}$. Furthermore, $Y^D\propto \tilde Y^D
v_{45}/\mpl$ and $v_{45}$ and $v_{\ov{16}}$ are the VEVs of the
$\mathbf{45_H}$ and $\mathbf{\ov{16}_H}$ fields. The SO(10) Higgs
fields comprise the SU(5) ones as $\mathbf{10}_\mathbf{H} \supset
\mathbf{5}_\mathbf{H} $ and $\mathbf{10_H^\prime} \supset
\mathbf{\ov{5}_H} $. The remaining components of the SO(10) Higgs
fields are assumed heavy with zero VEVs.  Finally, at the GUT scale
the SU(5) theory is broken to the MSSM and the MSSM Higgs fields $H_u
\subset \mathbf{5}_\mathbf{H}$ and $H_d \subset \mathbf{\ov{5}_H}$ 
couple to up- and down-type fermions, respectively.

The soft SUSY-breaking terms are assumed universal near the Planck
scale. The large Yukawa coupling in $Y^U$ now renormalises the squark
mass matrix. The renormalisation group (RG) flow down to $\mten$ (and to
$\mgut$) will keep its diagonal form (in the basis in which $Y^U$
and $Y^M$ are diagonal), but will split the mass of the third from
those of the first two generations. The diagonalisation of $Y^D$  
involves the rotation of $\Phi$ in \eq{w5} with $U_{\rm PMNS}$.
Since  $\Phi$ unifies the left-handed (s)leptons with righthanded
(s)quarks, the large atmospheric neutrino mixing angle will appear in 
the mixing of right-handed $\tilde b$ and $\tilde s$ squarks. 

We have performed a complete RG analysis of the CMM model and computed
its impact on \bbms. The key parameter in the analysis is the top
Yukawa coupling $y_t$, which drives the $\tilde b_R$--$\tilde s_R$
mixing effect. The fact that the RG evolution of this coupling is
governed by infrared quasi-fixed points in SO(10), SU(5) and the MSSM
allows us to place an upper bound on \bbms. At this point we remark
that our result corresponds to the Higgs sector specified above
(supplemented by an additional $\mathbf{16}$ to avoid unwanted D-term
breaking), which is minimal in its effect on the RG evolution of $y_t$
and leads to the weakest possible bound on \bbms. Changing e.g.\ the
last term in \eq{w10} into a dimension-4 coupling involving a
$\mathbf{\ov{126}_H}$ Higgs field while keeping the low-energy
parameters $m_t$ and $\tan \beta$ fixed will strengthen our bound on
\bbms\ further.

\section{The top Yukawa coupling}
\begin{figure}
  \centering
  \psfragscanon
  \psfrag{yt}{$y_t$}
  \psfrag{yt_c}{$y_t^c$}
  \psfrag{LogMuGev}{$\log_{10} \mu [\!\gev]$}
  \resizebox{0.45\textwidth}{!}{\includegraphics[]{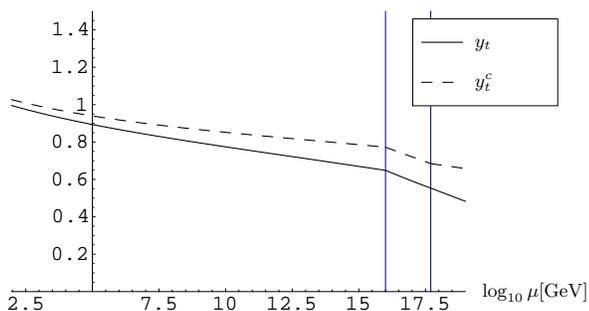}}
\caption{RG evolution of $y_t$. See text for explanation.
\label{fig:yt}}
\end{figure}

The large top Yukawa coupling $y_t$ suppresses the third-generation
squark (and slepton) masses through its RG effects and generates the
desired flavour-changing neutral $\tilde b_R$--$\tilde s_R$
transitions. In both SO(10) and SU(5) $y_t$ possesses an infrared
quasi-fixed point corresponding to a fixed point of the ratio $y_t/g$,
where $g$ is the gauge coupling. From the values of $m_t$ and $\tan
\beta$, which is the ratio of the VEVs of $H_u$ and $H_d$, one can
compute $y_t\propto m_t/\sin \beta$ at the electroweak scale and
evolve it up to $\mgut$, $\mten$ and the fundamental scale near
$\mpl$. The running above $\mten$ is much stronger than between
$\mgut$ and $\mten$ because the group-theoretical factors in
SO(10) are larger than in SU(5).  Next we consider the critical Yukawa
coupling $y_t^c$, which corresponds to the quasi-fixed point in
SO(10), i.e.\ $y_t^c/g$ is constant above $\mten$. $y_t^c$ is shown
as a dashed curve in Fig.~\ref{fig:yt}, with the two vertical lines
indicating the scales $\mgut$ and $\mten$. For $y_t<y_t^c$ one
consequently finds $y_t$ small at high energies. The solid line in
Fig.~\ref{fig:yt} illustrates this situation for the following input
parameters:
\vspace*{1mm}\\
\begin{tabular}{lllll}
\hline\noalign{\smallskip}
	$\mtpole$ & 
	$\alpha_s(M_Z)$ &
	$\alpha_2(M_Z)$ &
	$\alpha_1(M_Z)$ &
	$\tan\beta$ \\
\hline &&&&\\[-3mm]
	174 GeV
	& 0.121
	& 0.034
	& 0.017
	& 3 \\
\vspace*{-3mm} \\
	$m_\tq$ &
	$m_{\stp_1}$ &
	$m_{\stp_2}$ &
	$\theta_\stp$ &
	$m_\tg$ \\
\hline &&&&\\[-3mm]
	300 GeV
	& 200 GeV
	& 300 GeV
	& $\pi/6$
	& 400 GeV \\
\noalign{\smallskip}\hline
\end{tabular}\vspace*{1mm}
\\
We have computed all relevant RG coefficients in SO(10) and SU(5)
using the general result of \cite{mv}. For the RGE's above $\mgut$ we
work in the leading logarithmic approximation, but include
next-to-leading-order corrections in the MSSM. In this way we can
account for electroweak threshold corrections, which become relevant
if $y_t$ is close to $y_t^c$. For instance, $y_t$ depends on the
listed squark masses through these corrections. $\alpha_{1,2,s}$ are
the usual squared MSSM gauge couplings divided by $4\pi$,
$\theta_\stp$ is the stop mixing angle and $m_\tg$ is the gluino
mass. One finds $y_t$ raised to $y_t^c$ for $(m_t,\tan \beta) =
(180\gev, 3)$ or for $(m_t,\tan
\beta) = (184\gev, 4)$ with the remaining parameters unchanged.
For very small $\tan\beta < 2$, the fixed-point values can be exceeded
and the coupling typically becomes nonperturbative below the Planck
scale. In this case the model loses its predictivity. Since further
such low values of $\tan\beta$ are strongly disfavoured by LEP data,
we require $y_t\leq y_t^c$.

\section{\bbms}
The \bbms\ amplitude $M_{12}$ can be expressed in terms of Wilson
coefficients, which contain the short-dis\-tance information,
and matrix elements of local four-quark operators. The 
Standard Model contribution involves only a single operator:
\bea
	O_L &=& \bar s_L \gamma_{\mu} b_L \;
			\bar s_L \gamma^{\mu} b_L . \no
\eea
Its Wilson coefficient $C_L$ is computed from the box diagram
involving two W bosons and top quarks. In the CMM model a new operator
$O_R$ occurs, which is obtained from $O_L$ by replacing the
left-handed fields with right-handed ones. Since $\bra{B_s} O_L
\ket{\ov{B}_s}=\bra{B_s} O_R \ket{\ov{B}_s}$, no new hadronic matrix
elements are needed. Using the standard relativistic normalisation 
$\bra{B_s} B_s \rangle = 2 E V$ we can write
\bea
 M_{12} &=& \frac{G_F^2\, M_W^2 }{32  \pi^2 \, M_{B_s}} \, \lambda_t^2
	\, 
	  \lt( C_L + C_R \rt) \bra{B_s} O_L
	\ket{\ov{B}_s} . \label{m12}
\eea
Here $\lambda_t= V_{ts}^* V_{tb}$ comprises the CKM elements. We
define all Wilson coefficients and matrix elements at the
renomalisation scale $\mu =m_b$, at which the Standard Model
coefficient evaluates to $C_L=8.5$ \cite{bjw}.  The leading
contribution to \bbms\ in the CMM model arises from one-loop box 
diagrams with gluinos and squarks. The result is 
\be
	C_R = 
	\frac{\Lambda_3^2}{\lambda_t^2} 
        \frac{8 \pi^2 \alpha_s^2 (m_{\tilde g}) }{G_F^2 M_W^2 m_{\tilde g}^2 }
	\left[ \frac{\alpha_s(m_{\tilde g})}{\alpha_s(m_b)}\right]^{6/23}
	S^{(\tg)} .
\ee
Here 
\be
	|\Lambda_3| = |U_{\mu 3}| |U_{\tau 3}| \approx \frac{1}{2}
		\label{eq:Lam3}
\ee
is the relevant combination of mixing matrix elements in the right-handed
sdown sector, and
\bea
	S^{(\tg)} &=& \, \frac{11}{18} \lt[ 
	G(x_{\tilde b},x_{\tilde b}) +
	G(x_{\tilde s},x_{\tilde s}) - 2 
	G(x_{\tilde b},x_{\tilde s}) \rt] \nn
	&& - \frac{2}{9} \lt[ 
	F(x_{\tilde b},x_{\tilde b}) +
	F(x_{\tilde s},x_{\tilde s}) - 2 
	F(x_{\tilde b},x_{\tilde s}) \rt]
\eea
with $x_{\tilde q}=m_{\tilde q}^2/m_{\tilde g}^2$ and $\tilde s,\tilde
b$ denoting the right-handed squarks of the second and third
generations. The functions $F$ and $G$ are defined
in~\cite{Bertolini:1990if}.  Note the twofold enhancement of $C_R$ due
to the large atmospheric mixing and the large strong coupling
constant. This is, however, partially offset by a smaller loop function
$S^{(\tg)}$. The neutral $\mathrm{B_s}$-meson mass difference is given
by
\be
	\Delta M_{B_s} = 2 | M_{12} |.
\ee
The phase of $M_{12}$ is responsible for mixing-induced CP
violation. Note that the phase of $\Lambda_3$ is undetermined, so that
there is potentially large, but not predictable, CP violation in decay
modes like $B_s \to \psi \phi, \psi \eta^{(\prime)}$. A measurement of
$\Delta M_{B_s}$ and the mixing-induced CP asymmetries in one of these
decays will allow to determine both magnitude and phase of $M_{12}$
and therefore of $C_R$. If the $\mathrm{B_s}$ oscillations are too rapid, these
asymmetries cannot be measured. Then still the width difference in the
$B_s$ system can be used to determine $\cos (\arg M_{12})$ \cite{dfn}.

The input parameters for our analysis are $y_t$, $m_{\tg}$ and the
average squark mass $m_{\tilde q}$ and the trilinear term $a_d$ of the
first generation, all defined at the electroweak scale. They are
evolved to the GUT scale and the SU(5) and SO(10) GUT parameters are
determined. All other parameters are then determined from the RG flow
back down to the electroweak scale. The flavour-changing effects in
the CMM model grow with the sizes of the universal soft sfermion mass
$m_0$ and the universal trilinear term $a_0$ at the Planck scale. 
However, a larger $m_0$ also implies larger squark masses leading to a
suppression of the gluino box function $S^{(\tg)}$. The effect on
\bbms\ is maximal for values of $m_{\tilde q}$ around 800
\gev. Furthermore $a_0$ and $m_0$, which for a given gluino mass 
determine the sfermion mass spectrum, are constrained by the lower
bounds on these masses.  We have further checked that no charge- or
colour-breaking vacuum occurs.  Finally the effect decreases with
increasing $m_{\tg}$, so that the maximal effect is found for the
experimental lower bound $m_{\tg}=195 \gev$.

\begin{figure}
  \centering
  \psfragscanon
  \psfrag{Re2M12}{$\Re\left(2 M_{12}\right)$}
  \psfrag{Im2M12}{$\Im\left(2 M_{12}\right)$}

  \resizebox{0.45\textwidth}{!}{\includegraphics[]{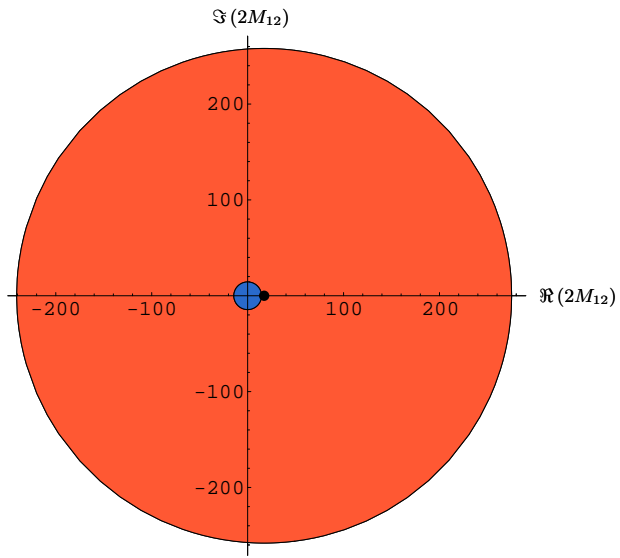}}
\caption{Allowed region in the complex $2 M_{12}$ plane. The axis units are
inverse picoseconds.} 
\label{fig:m12_plane}
\end{figure}

The allowed area in the complex $M_{12}$ plane for the CMM model is
depicted in Fig.~\ref{fig:m12_plane}. For simplicity we neglect the
30\% uncertainty from current lattice QCD determinations of $\bra{B_s}
O_L \ket{\ov{B}_s}$. Fixing the value of this matrix elements to the
central value quoted in \cite{Lellouch:2002nj} results in a Standard
Model prediction of $17.2\; \mbox{ps}^{-1}$. The small black circle in
Fig.~\ref{fig:m12_plane} indicates this value and the large
red disk denotes the range covered by the CMM model.
The blue circle centered at the origin is the region excluded
by the 95 \% CL experimental lower bound~\cite{Battaglia:2003in}
\be
	\Delta M_{B_s} > 14.4\,\mbox{ps}^{-1} .
\ee
We see that in total the mass difference can exceed its Standard-Model
prediction by a factor of 16.

In conclusion we have perfomed a RG analysis for the CMM model
\cite{cmm} and computed the impact on \bbms. Our work complements and
improves previous GUT-inspired analyses (see e.g.\ \cite{hlmp}), which
supplement the MSSM with minimal flavour violation by $\tilde
b_R$--$\tilde s_R$ mixing at the electroweak scale.

%

\section*{Acknowledgments}
This work is supported by the German BMBF under contract No.~05HT1WOA3.
Fermilab is operated by Universities Research Association Inc.\ under
Contract No.~DE-AC02-76CH03000 with the United States Department of
Energy.

\newpage
{\small
\emph{Disclaimer:}\\
This report was prepared as an account of work sponsored by an agency
of the United States Government. Neither the United States Government
nor any agency thereof, nor any of their employees, makes any
warranty, expressed or implied, or assumes any legal liability or
responsibility for the accuracy, completeness, or usefulness of any
information, apparatus, product, or process disclosed, or represents
that its use would not infringe privately owned rights. Reference
herein to any specific commercial product, process, or service by
trade name, trademark, manufacturer, or otherwise, does not
necessarily constitute or imply its endorsement, recommendation, or
favoring by the United States Government or any agency thereof. The
views and opinions of authors expressed herein do not necessarily
state or reflect those of the United States Government or any agency
thereof.  Distribution Approved for public release; further
dissemination unlimited.

\emph{Copyright notification:}\\
Notice: This manuscript has been authored by Universities Research
Association, Inc. under contract 
No.~{\scriptsize DE-AC02-76CH03000} with the
U.S. Department of Energy. The United States Government retains and
the publisher, by accepting the article for publication, acknowledges
that the United States Government retains a nonexclusive, paid-up,
irrevocable, worldwide license to publish or reproduce the published
form of this manuscript, or allow others to do so, for United States
Government purposes.
}

\end{document}